\documentclass[onecolumn,12pt,journal,draftclsnofoot,journalpaper]{IEEEtran}

\usepackage{amssymb}
\usepackage{amsmath}
\usepackage{amsthm}
\usepackage{amsfonts}
\usepackage{mathrsfs}
\usepackage{float}
\usepackage{graphicx}
\usepackage{epsfig,graphics,subfigure,psfrag}
\usepackage{pdfpages}
\usepackage{enumerate}
\usepackage{engord}
\usepackage{cite}
\usepackage{verbatim}
\usepackage[ruled,linesnumbered]{algorithm2e}

\begin{document}

\newtheorem{definition}{\bf Definition}
\newtheorem{theorem}{\bf Theorem}
\newtheorem{lemma}{\bf Lamma}
\newtheorem{corollary}{\bf Corollary}
\newtheorem{proposition}{\bf Proposition}

\title{UAV Offloading: Spectrum Trading Contract \\ Design for UAV Assisted 5G Networks}
\author{
\IEEEauthorblockN{
\normalsize{Zhiwen Hu},
\normalsize{Zijie Zheng},
\normalsize{Lingyang Song},
\normalsize{Tao Wang},
and
\normalsize{Xiaoming Li}
 \\}
\IEEEauthorblockA{\normalsize{School of Electronics Engineering and Computer Science, Peking University, Beijing, China\\
Email: \{zhiwen.hu, zijie.zheng, lingyang.song, wangtao, lxm\}@pku.edu.cn} \\
}
}

\maketitle

\begin{abstract}

Unmanned Aerial Vehicle (UAV) has been recognized as a promising way to assist future wireless communications due to its high flexibility of deployment and scheduling.
In this paper, we focus on temporarily deployed UAVs that provide downlink data offloading in some regions under a macro base station~(MBS).
Since the manager of the MBS and the operators of the UAVs could be of different interest groups, we formulate the corresponding spectrum trading problem by means of contract theory, where the manager of the MBS has to design an optimal contract to maximize its own revenue.
Such contract comprises a set of bandwidth options and corresponding prices, and each UAV operator only chooses the most profitable one from all the options in the whole contract.
We analytically derive the optimal pricing strategy based on fixed bandwidth assignment, and then propose a dynamic programming algorithm to calculate the optimal bandwidth assignment in polynomial time.
By simulations, we compare the outcome of the MBS optimal contract with that of a social optimal one, and find that a selfish MBS manager sells less bandwidth to the UAV operators.

\end{abstract}


\newpage
\section{Introduction}\label{sec_Introduction}

The rapid development of wireless communication enabled small-scale unmanned aerial vehicles (UAVs) has created a variate of civil applications~\cite{bib_CivilApplicationsOfUAV}, from cargo delivery\cite{bib_CargoDelivery} and remote sensing~\cite{bib_RemoteSensing} to data relaying~\cite{bib_AirborneCommunicationNetworks} and connectivity maintenance~\cite{bib_FlyingAdHocNetwork,bib_ThroughputMaximizationRelay}.
From the aspect of wireless communications, one major advantage of utilizing UAVs is their high probability of keeping line-of-sight (LoS) signals with other communication nodes, alleviating the problem brought by severe shadowing in urban or mountainous terrain~\cite{bib_RadioChannelModeling,bib_AirGroundChannel}.
Different from high-altitude platforms which are designed for long-term assignment above tens of kilometers height~\cite{bib_HighAltitudePlatform}, small-scale UAVs within only hundreds of meters off the ground can be deployed more quickly.
In addition, the properties like low-cost, high flexibility and ease of scheduling also make small-scale UAVs a favorable choice in civil usages, in spite of their disadvantages such as low battery capacity~\cite{bib_WirelessCommunicationsWithUAV}.

One of the major problems in the UAV assisted wireless communications is to optimally deploy UAVs, in which way mobile users can be better served~\cite{bib_WirelessCommunicationsWithUAV}.
Many studies have been done to deal with this problem from distinctive viewpoints with respect to different objectives and constraints~\cite{bib_OptimalAltitudeForCoverage,bib_DroneSmallCell,bib_3DPlacementOfUAV,bib_UAVD2D,bib_DownlinkCoverageProbability,bib_PlacementOptimization,bib_EfficientDeploymentOfMultipleUAVs,bib_OptimalPowerEfficientDeployment,bib_InterferenceAwarePositioning,bib_UAVAsistedPublicSafety,bib_UAVAsistedCapacityEnhance,bib_CellAssociation,bib_ResourceAllocationMovingSmallCell}.
Among them, the works in~\cite{bib_OptimalAltitudeForCoverage,bib_DroneSmallCell,bib_3DPlacementOfUAV,bib_UAVD2D} considered the scenario consisting only one UAV to provide with coverage, the works in~\cite{bib_DownlinkCoverageProbability,bib_PlacementOptimization,bib_EfficientDeploymentOfMultipleUAVs,bib_OptimalPowerEfficientDeployment} took into account multiple UAVs to providing better services by joint coverage, and the works in~\cite{bib_InterferenceAwarePositioning,bib_UAVAsistedPublicSafety,bib_UAVAsistedCapacityEnhance,bib_CellAssociation,bib_ResourceAllocationMovingSmallCell} studied the coexistence of base stations (BSs) and multi-UAVs, where data offloading becomes a major problem.

To be specific, in~\cite{bib_OptimalAltitudeForCoverage}, the optimal height of a single UAV was deduced to maximize the coverage radius.
The authors of~\cite{bib_DroneSmallCell} minimized the transmission power of the UAV with fixed coverage radius.
The problem of maximizing the number of users that covered by one UAV is studied in~\cite{bib_3DPlacementOfUAV}.
And the authors of~\cite{bib_UAVD2D} further took into account the interference from device-to-device (D2D) users.
For multiple UAVs, the coverage probability of a ground user was derived in~\cite{bib_DownlinkCoverageProbability} .
The work in~\cite{bib_PlacementOptimization} proposed a solution to minimize the number of UAVs to cover all the users.
The authors of~\cite{bib_EfficientDeploymentOfMultipleUAVs} studied the deployment of multiple UAVs to achieve largest total coverage area.
And in~\cite{bib_OptimalPowerEfficientDeployment}, the total transmission power of UAVs was minimized while the data rate for each user was guaranteed.
With the consideration of BSs in the scenario, the gain of deploying additional UAVs for offloading was discussed in~\cite{bib_InterferenceAwarePositioning,bib_UAVAsistedPublicSafety,bib_UAVAsistedCapacityEnhance}.
The authors of~\cite{bib_CellAssociation} focused on the optimal cell partition strategy to minimize average delay of the users in a cellular network with multiple UAVs.
In~\cite{bib_ResourceAllocationMovingSmallCell}, the optimal resource allocation was presented, where one MBS, multiple small-cell base stations (SBSs) and multiple UAVs are involved.

Although UAV coverage and offloading problems have been widely discussed, few existing studies consider the situation where UAV operators could be selfish individuals with different objectives~\cite{bib_EconomicMotivationsOfUAV}.
For instance, the venue owners and scenic area managers may want to temporarily deploy their own UAVs to better serve their visitors, due to the temporarily increased number of mobile users or the inconvenience of installing SBSs in remote areas~\cite{bib_TemporaryEvents}.
In such cases, the deployment of multiple UAVs depends on each UAV operator, and the solution is not likely to be optimal as calculated by centralized algorithms.
In addition, the wireless channel allocation becomes a more critical problem since the bandwidth that the UAVs used to serve mobile users has to be explicitly authorized by the MBS manager.
Therefore, further studies need to be done with respect to selfish UAV operators in UAV assisted offloading cellular networks.

In this paper, we focus on the scenario with one MBS that managed by the MBS manager, and multiple SBS-enabled UAVs that owned by different UAV operators.
To enable downlink transmissions of the UAVs, each UAV operator has to buy a certain amount of bandwidth that authorized by the MBS manager.
However, the total usable bandwidth of the MBS is limited, and selling part of the total bandwidth to the UAVs may harm the capacity of the MBS.
Therefore, payments to the MBS manager should be made by UAV operators.
Here, contract theory~\cite{bib_ContractTheory} can be applied as a tool to analyze the optimal contract that the MBS manager will design to maximize its revenue.
Specifically, such contract comprises a set of bandwidth options and corresponding prices.
Since each UAV operator only chooses the most profitable option from the whole contract, the MBS manager has to guarantee that the contract is feasible, i.e., the option that a UAV operator chooses from the contract is exactly the one that designed for it.

The main contributions of our work are listed as below:
\begin{enumerate}
\item We formulate the optimal contract design problem where the selfish MBS manager has to decide the number of channels and the amount of price that designed for each type of selfish UAV operator in order to realize data offloading.
\item We analytically deduce the optimal pricing strategy and propose our dynamic programming algorithm to achieve the optimal bandwidth allocation efficiently.
\item We reveal some significant insights based on the simulation results, e.g., the selfish MBS manager sells less bandwidth to the UAV operators compared with a social optimal result.
\end{enumerate}

The rest of our paper is organized as follows.
Section II presents our system model and formulates the optimal contract design problem.
Section III theoretically deduces the optimal solution and provides our dynamic programming algorithm.
Section IV focuses on the height of the UAVs and discuss its impact on the revenue of the MBS manager.
Section V shows the simulation results of the optimal contract.
Finally, we conclude our paper in Section VI.

\section{System Model}\label{sec_SystemModel}

\vspace{-5mm}
\begin{figure}[!thp]
\centering
\includegraphics[width=3.5in]{{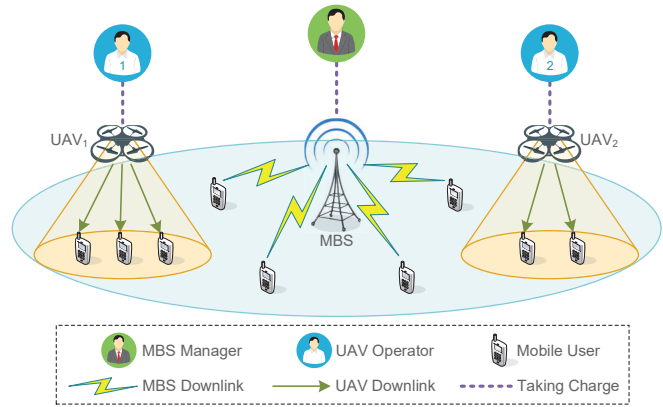}}
\vspace{-5mm}
\caption{The system model of UAV Assisted Offloading in Cellular Networks with $1$ MBS manager and multiple UAV operators.}\label{fig_SystemModel}
\vspace{-5mm}
\end{figure}

We consider a scenario with one MBS that run by the MBS manager, and $N$ UAVs that run by $N$ different UAV operators, as shown in Fig.~\ref{fig_SystemModel}.
The UAVs that deployed by the UAV operators are assumed to serve mobile users with licensed spectrum\footnote{The wireless backhaul connections between the UAVs and the MBS, however, are assumed to follow the standard MBS-SBS communication regulations, and thus is not our major concern in this paper.}, and to be deployed at a unified altitude $H$ (designated by the MBS manager) and at different fixed horizontal locations.
Each UAV operator first has to buy a certain amount of bandwidth from the MBS manager, where the utility and the cost of each individual should be addressed.

In the rest part of this section, we first present the wireless downlink model of the MBS and the UAVs, then introduce the utility of the UAV operators as well as the cost of the MBS manager, and finally formulate the contract design problem.

\vspace{-4mm}
\subsection{Wireless Downlink Model}

The air-to-ground wireless channel between a UAV and a mobile user mainly consists of two parts, which are the Line-of-Sight (LoS) component and the None-Line-of-Sight (NLoS) component~\cite{bib_AirGroundChannel}.
Based on the study in~\cite{bib_OptimalAltitudeForCoverage}, the probability of LoS for a user with elevation angle $\theta$ (in degree) to a specific UAV is given by $P_{LoS}(\theta)=\dfrac{1}{1+a\exp{\big(-b[\theta-a]\big)}}$,
where $a$ and $b$ are the parameters that depend on the specific terrain (like urban, rural, etc.).

Based on $P_{LoS}$, the average pathloss from the UAV to the user can be given by (in dB):
\begin{equation}\label{eqn_PathLoss}
\left\{
\begin{array}{ll}
 \!\!\!\! & \overline{L}_{UAV}(\theta,d) = P_{LoS}(\theta)\cdot L_{LoS}(d)+\big[1-P_{LoS}(\theta)\big]\cdot L_{NLoS}(d), \\
 \!\!\!\! & L_{LoS}(d) = 20\log{(4\pi fd/c)}+\eta_{LoS}, \\
 \!\!\!\! & L_{NLoS}(d) = 20\log{(4\pi fd/c)}+\eta_{NLoS},
\end{array}
\right.
\end{equation}
where $c$ is the speed of light, $d$ is the distance between the UAV and the user, and $f$ is the frequency of the channel.
$L_{LoS}(d)$ and $L_{NLoS}(d)$ are the pathloss of the LoS component and the pathloss of the NLoS component, respectively.
$\eta_{LoS}$, $\eta_{NLoS}$ are the average additional loss that depends on the environment.
In contrast to the UAV-to-user wireless channel, the MBS-to-user channels are considered as NLoS only, which gives us the average pathloss as:
\begin{equation}
\overline{L}_{MBS}(d)=20\log{(4\pi fd/c)}+\eta_{NLoS}.
\end{equation}
For simplicity, we assume that different channels has similar $f$ and the difference can be ignored.

To see the signal quality that each user could experience, we use $\gamma_{MBS}(d)$ to denote the Signal-to-Noise Ratio (SNR) for MSB users at the distance $d$ from the MBS.
And we have $\gamma_{MBS}(d)=\big[P_{MBS}-\overline{L}_{MBS}(d)\big]/N_0$, where $P_{MBS}$ is the transmission power of the MBS and $N_0$ is the power of background noise.
Similarly, we use $\gamma_{UAV}(d,\theta)$ to denote the SNR for the UAV users with elevation angle $\theta$ and distance $d$ from a certain UAV, given as $\gamma_{UAV}(d,\theta)=\big[P_{UAV}-\overline{L}_{UAV}(d,\theta)\big]/N_0$, where $P_{UAV}$ is the transmission power of the UAV.

It is also assumed that each user can automatically choose among the MBS and the UAVs to obtain the best SNR.
Therefore, it is necessary to find out in which region a certain UAV is able to provide better SNR than the others (including the MBS and the other UAVs).
We denote the region where UAV$_n$ provides better SNR as UAV$_n$'s effective offloading region, denoted by $\Omega_n$.
Let $\Lambda_n(\boldsymbol{x})$ denote the boolean variable that indicates whether the user with the location $\boldsymbol{x}$ (a two-dimensional vector) is in $\Omega_n$.
Thus the area of $\Omega_n$ can be calculated by $S_n=\int \Lambda_i(\boldsymbol{x}) d\boldsymbol{x}$.

\vspace{-4mm}
\subsection{The Utility of the UAV operators}

Each mobile user in an effective offloading region is assumed to access to the UAV randomly.
We call the density of the users in $\Omega_n$ that want to connect to UAV$_n$ at any instant as the ``active user density" of UAV$_n$, denoted by $\sigma_n$.
And the corresponding concept, ``active user number" $\varepsilon_n$, describing the number of users that want to set up connects with UAV$_n$ in any moment, is given by $\varepsilon_n=S_n \sigma_n$.
We assume that $\sigma_n$ obeys Poisson distribution with mean value of $\rho_n$.
Considering a given location of UAV$_n$ (which makes $S_n$ a constant), $\varepsilon_n$ also becomes an Poisson-distribution variable, with mean value $\mu_n=S_N \rho_n$.
Based on $\mu_n$, we can classify the UAVs into multiple types.
Specifically, we refer to UAV$_n$ as a $\lambda$-type UAV if $\mu_n=\lambda$, which means that there are averagely $\lambda$ users connecting to UAV$_n$ at any instant.
The number of $\lambda$-type UAVs is denoted by $N_\lambda$, where $\sum_{\lambda} N_\lambda = N$.
For writing simplicity, we use random variable $X_\lambda$ (instead of $\varepsilon_n$) to denote the active user number of a $\lambda$-type UAV.
The probability of $X_\lambda=k$ is given by
\begin{equation}
P(X_{\lambda}=k) = \frac{(\lambda)^k}{k!} e^{-\lambda}, \quad\quad  k=0,1,2,\cdots
\end{equation}

Without the loss of generality, we assume that each mobile user connecting to a UAV (or the MBS) is allocated with one channel with fixed bandwidth $B$, in a frequency division pattern.
Due to the variation of the active user number, there is always a probability that an UAV fails to serve the current active users.
Therefore, the more channels are being obtained, the more utility the UAV can achieve.
The utility function of obtaining $w$ channels for a $\lambda$-type UAV is denoted by $U(\lambda,w)$.
Since the utility of obtaining no channels is $0$, we have
\begin{equation}\label{eqn_UAVUtilityDefinition1}
U\big(\lambda,w\big)=0, \quad\quad w=0.
\end{equation}
The marginal utility of obtaining another channel depends on the probability that the current number of channels is not sufficient for random user requests.
Thus we have
\begin{equation}\label{eqn_UAVUtilityDefinition2}
U\big(\lambda,w\big)=U\big(\lambda,(w-1)\big)+P(X_{\lambda} \ge w), \quad\quad w \ge 1.
\end{equation}
Based on~(\ref{eqn_UAVUtilityDefinition1}) and~(\ref{eqn_UAVUtilityDefinition2}), we can derive the general term of the $\lambda$-type UAV's utility as
\begin{equation}
U\big(\lambda,w\big)=\sum\limits_{k=1}^{k=w} P(X_{\lambda} \ge k), \quad\quad w\ge 1.
\end{equation}

\vspace{-4mm}
\subsection{Cost of the MBS manager}

It is assumed that the MBS will not reuse the spectrum that is already sold, which implies the MBS manager suffers a certain degree of loss as it sells the spectrum to UAV operators.
The active user number of the MBS is also assumed to follow the Poisson distribution.
We denote this random variable as $X_{BS}$ and the mean value of it as $\lambda_{BS}$.
Therefore we have
\begin{equation}
P(X_{BS}=k) = \frac{(\lambda_{BS})^k}{k!} e^{-\lambda_{BS}}, \quad\quad  k=0,1,2,\cdots
\end{equation}
The total number of channels of the MBS is denoted by $M$, $M \in \mathbb{Z}^{+}$.
Just like the situation of UAVs, there is also a utility of a certain number of channels for the MBS manager, $U_{BS}(m)$, representing the average number of users that $m$ channels can serve, given as
\begin{equation}\label{eqn_UtilityDefinition1}
U_{BS}\big(m\big)=0, \quad\quad m=0,
\end{equation}
\begin{equation}\label{eqn_UtilityDefinition2}
U_{BS}\big(m\big)=U_{BS}\big(m-1\big)+P(X_{BS} \ge m), \quad\quad m\ge 1.
\end{equation}
Based on the utility of the MBS manager, we define the cost function $C\big(m\big)$ as the utility loss of reducing the number of channels from $M$ to $M-m$, given as
\begin{equation}\label{eqn_Cost}
C\big(m\big)=U_{BS}\big(M\big)-U_{BS}\big(M-m\big)=\sum\limits_{k=M-m+1}^{M} P(X_{BS} \ge k).
\end{equation}

\vspace{-4mm}
\subsection{Contract Formulation}

Since different types of UAVs have different demands, the MBS manager has to design a contract which contains a set of ``quality-price" options for all the UAV operators, denoted by $\Big\{\big(w(\lambda), p(\lambda)\big) \,\big|\, \forall \lambda \in \Lambda\Big\}$.
In this contract, $w(\lambda)$ is the number of channels that designed to sell to a $\lambda$-type UAV operator, and $p(\lambda)$ is the corresponding price designed to be charged.
Each $\big(w(\lambda), p(\lambda)\big)$ pair can be seen as a commodity with quality $w(\lambda)$ at price $p(\lambda)$.

However, each UAV operator is expected to choose the one that maximize its own profit according to the whole contract.
The contract is feasible if and only if any $\lambda$-type UAV operator prefers the commodity $\big(w(\lambda), p(\lambda)\big)$ to all the others~\cite{bib_SpectrumContract}.
And to achieve this, the first requrrement is the incentive compatible (IC) condition, implying that the commodity designed for a $\lambda$-type UAV operator in the contract is indeed the best one for it, given by
\begin{equation}\label{eqn_IC}
U\big(\lambda,w(\lambda)\big) -p(\lambda) \ge U\big(\lambda,w(\lambda^\prime)\big) -p(\lambda^\prime), \quad\quad \forall \lambda^\prime \ne \lambda.
\end{equation}
The second requirement is the individual rational (IR) condition, meaning that the $\lambda$-type UAV operator will not buy any channels if all of the options bring negative profits, given by
\begin{equation}\label{eqn_IR}
U\big(\lambda,w(\lambda)\big) -p(\lambda) \ge U\big(\lambda,0\big) -0=0,
\end{equation}
where $U\big(\lambda,0\big) -0$ implies an ``empty commodity'' in the contract, which has no utility and no price to be charged.
To put it simple, a feasible contract has to satisfy the IC constraint and the IR constraint, and any contract that satisfies the IC and IR constraints is feasible.

For the MBS manager, the overall revenue brought by the contract $\{w(\lambda), p(\lambda) \,|\, \forall \lambda \in \Lambda\}$ is
\begin{equation}\label{eqn_Revenue}
R=\sum\limits_{\lambda \in \Lambda} \Big(N_\lambda \cdot p(\lambda) \Big) -  C\Big(\sum\limits_{\lambda \in \Lambda} N_\lambda \cdot w(\lambda)\Big),
\end{equation}
where $N_\lambda \cdot p(\lambda) $ is the total payment obtained from $\lambda$-type UAV operators, and $\sum\limits_{\lambda \in \Lambda} N_\lambda \cdot w(\lambda)$ is the total number of channels that being sold.
The objective of the MBS manager is to design proper $w(\lambda)$ and $p(\lambda)$ for any given $\lambda \in \Lambda$, in which way it can maximize its own revenue with the pre-consideration of each UAV operator's behavior, given as
\begin{equation}\label{eqn_ObjectiveFunction}
\begin{split}
\hat{R} = &  \max\limits_{\{w(\lambda)\} , \{p(\lambda)\}} \sum\limits_{\lambda \in \Lambda} \Big(N_\lambda \cdot p(\lambda) \Big) -  C\Big(\sum\limits_{\lambda \in \Lambda} N_\lambda \cdot w(\lambda)\Big), \\
 s.t. \quad & U\big(\lambda,w(\lambda)\big) -p(\lambda) \ge U\big(\lambda,w(\lambda^\prime)\big) -p(\lambda^\prime) \ge 0, \quad\quad \forall \lambda, \lambda^\prime \in \Lambda \textrm{ and } \lambda^\prime \ne \lambda, \\
 & U\big(\lambda,w(\lambda)\big) -p(\lambda) \ge 0, \quad\quad\quad\quad\quad\quad\quad\quad\quad\quad\quad\,\,\, \forall \lambda, \lambda^\prime \in \Lambda \textrm{ and } \lambda^\prime \ne \lambda, \\
 & p(\lambda) \ge 0, \quad w(\lambda)=0,1,2\cdots \quad\quad\quad\quad\quad\quad\quad\quad\quad\,\, \forall \lambda \in \Lambda, \\
 & \sum\limits_{\lambda \in \Lambda} N_\lambda \cdot w(\lambda)\le M,
\end{split}
\end{equation}
where the first two constraints represent the IC and the IR, and the last one indicates the limited number of channels possessed by the MBS.
In the rest part of our paper, the quality assignment $w(\lambda)$, and the pricing strategy $p(\lambda)$, are the two most basic concerns.
In addition, we call the contract that optimizes the problem in~(\ref{eqn_ObjectiveFunction}) as the ``MBS optimal contract".

\section{Optimal Contract Design}\label{sec_Theoretical}

In this section, we exploit some basic properties of our problem in Section~\ref{sec_Design_Properties}.
By utilizing these properties, we provide the optimal pricing strategy based on the fixed quality assignment in Section~\ref{sec_Design_Pricing}. Next, we analyze and transform the optimal quality assignment problem in Section~\ref{sec_Design_Quality}, in which way it can be solved by the proposed dynamic programming algorithm given in Section~\ref{sec_Design_Algorithm}.
And finally we discuss the social optimal contract in Section~\ref{sec_Design_Social}.

To facilitate the writing of our following discussions, we put all the types $\{ \lambda \}$ in the ascending order, given by $\{ \lambda_1, \cdots \lambda_t, \cdots \lambda_T \}$ where $1\le t\le T$ and $\lambda_{t_1} < \lambda_{t_2}$ if $t_1 < t_2$.
Note that, in this case we call $\lambda_{t_1}$ as a ``lower type" and $\lambda_{t_2}$ as a ``higher type".
In addition, we also simplify $N_{\lambda_t}$ as $N_t$, $w(\lambda_t)$ as $w_t$ and $p(\lambda_t)$ as $p_t$ in the discussions below.

\vspace{-4mm}
\subsection{Basic Properties}\label{sec_Design_Properties}

\begin{figure}[!thp]
\centering
\includegraphics[width=5.0in]{{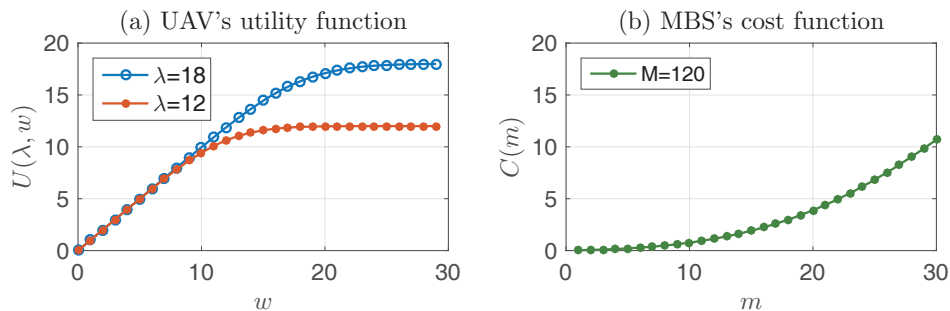}}
\vspace{-5mm}
\caption{An illustration of the profiles of the UAV's utility function and the MBS's cost function.}\label{fig_UtilityAndCost}
\vspace{-5mm}
\end{figure}

We first take a look at the utility function $U(\lambda,w)$.
As given in Fig.~\ref{fig_UtilityAndCost}, we can see that the marginal utility for a UAV operator decreases as the quality $w$ gets larger.
In addition, a UAV with higher type ($\lambda\!=\!18$) consider a certain bandwidth quality $w$ more valuable than a lower type ($\lambda\!=\!12$).
To prove the above observations, we first have to provide a more basic conclusion with respect to a property of Poisson distribution, on which the utility function is defined.

\vspace{-3mm}
\begin{lemma}\label{pro_Poisson}
Given that $X_\lambda$ and $X_{\lambda^\prime}$ are two Poisson distribution random variables with mean values $\lambda$ and $\lambda^\prime$ respectively, if $\lambda>\lambda^\prime>0$, then $P(X_{\lambda} \ge k) > P(X_{\lambda^\prime} \ge k)$ for any $k\in\mathbb{Z}^+$.
\end{lemma}
\vspace{-3mm}

The proof of Lemma~\ref{pro_Poisson} can be found in Appendix~\ref{sec_Appendix_Lem1}.
This lemma is particularly singled out since it is used in many of the following propositions.

\vspace{-3mm}
\begin{proposition}\label{pro_Monotonicity}
The utility function $U(\lambda,w)$ monotonously increases with the type $\lambda$ and the quality $w$, where $\lambda>0$ and $w\in\mathbb{N}$.
In addition, the marginal increase of $U(\lambda,w)$ with respect to $w$ gets smaller as $w$ increases.
\end{proposition}
\vspace{-3mm}

The proof of Proposition~\ref{pro_Monotonicity} is provided in Appendix~\ref{sec_Appendix_Pro1}.
This proposition provides a basic property for us to design the optimal contract in the rest of our paper.
From Fig.~\ref{fig_UtilityAndCost} (a) we can also notice that $U(\lambda,w)$ converges to a fixed value as $w\rightarrow\infty$, given by
\begin{equation}
\lim\limits_{w\rightarrow\infty} U\big(\lambda,w\big)=\sum\limits_{k=1}^{\infty} P(X_{\lambda}\ge k)=\sum\limits_{k=1}^{\infty}k\cdot \frac{(\lambda)^k}{k!} e^{-\lambda}=\sum\limits_{k=0}^{\infty}k\cdot \frac{(\lambda)^k}{k!} e^{-\lambda}=\lambda.
\end{equation}
Intuitively, the utility of a $\lambda$-type UAV cannot exceed its active user number even if excessive channels are allocated.
Such conclusion provides another way to comprehend the meaning of $U(\lambda,w)$, that is, the number of effectively served users of the UAV.

Based on Lemma~\ref{pro_Poisson} and Proposition~\ref{pro_Monotonicity}, we exploit another important property of $U(\lambda,w)$, which says that a certain amount of quality improvement is more attractive to a higher type UAV than a lower type UAV.
This property can be referred to as the ``increasing preference (IP) property", and we write it as the following proposition:

\vspace{-3mm}
\begin{proposition}\label{pro_IP} \textbf{(IP property) :}
For any UAV types $\lambda > \lambda^\prime > 0$ and channel qualities $w > w^\prime \ge 0$, the following inequality holds: $U(\lambda,w)-U(\lambda,w^\prime)>U(\lambda^\prime,w)-U(\lambda^\prime,w^\prime)$.
\end{proposition}
\vspace{-3mm}

The proof of IP property is given in Appendix~\ref{sec_Appendix_Pro2}.
With the help of this property, we are able to deduce the best pricing strategy in the next subsection.

\vspace{-4mm}
\subsection{Optimal Pricing Strategy}\label{sec_Design_Pricing}

In this subsection, we use fixed quality assignment $\{w_t\}$ to analytically deduce the optimal pricing strategy $\{p_t\}$.

Based on the previous work on contract theory (such as in~\cite{bib_SpectrumContract}), the IC \& IR constraints and the IP property of the utility function in a contract design problem can directly lead to the conclusion as below:

\vspace{-3mm}
\begin{proposition}\label{pro_PreviousWork}
For the contract $\big\{ (w_t, p_t) \big\}$ with the IC \& IR constraints and the IP property, the following statements are simultaneously satisfied:
\begin{itemize}
\item{The relation of types and qualities: $\lambda_{i} <\lambda_{j} \, \Longrightarrow \, w_{i} \le w_{j}$.}
\item{The relation of qualities and prices: $w_{i} <w_{j} \, \Longleftrightarrow \, p_{i} <p_{j}$.}
\end{itemize}
\end{proposition}
\vspace{-3mm}

This conclusion contains basic properties of a feasible contract.
It indicates that a higher price has to be associated with a higher quality, and a higher quality means higher price should be charged.
Although different qualities are not allowed to be associated with the same price, it is still possible that different types of UAVs are assigned with the same channel quality and thus the same price.

\vspace{-3mm}
\begin{lemma}\label{pro_Conditions}
For the contract $\big\{ (w_t, p_t) \big\}$ with the IC \& IR constraints and the IP property, the folowing three conditions are the necessary conditions and sufficient conditions to determine a feasible pricing:
\begin{itemize}
\item{$0 \le w_1 \le w_2 \le \cdots \le w_T$},
\item{$0\le p_1 \le U(\lambda_1,w_1)$},
\item{$p_{k-1}+A \le p_k \le p_{k-1} +B$, \, for $k=2,3,\cdots,T$, \\ where $A=\big[U(\lambda_{k-1},w_{k})-U(\lambda_{k-1},w_{k-1})\big] $ and $B=\big[U(\lambda_k,w_k)-U(\lambda_k,w_{k-1})\big]$.}
\end{itemize}
\end{lemma}
\vspace{-3mm}

The proof of Lemma~\ref{pro_Conditions} is given in Appendix~\ref{sec_Appendix_Lem1}.
It provides an important guideline to design the prices for different types of UAVs.
It implies that with fixed quality assignment $\{w_t\}$, the proper scope of the price $p_k$ depends on the value of $p_{k-1}$.

In the following, we provide the optimal pricing strategy of the MBS manager with fixed quality assignment $\{w_t\}$.
Here we call $\{w_t\}$ a feasible quality assignment if $w_1 \le w_2 \le \cdots \le w_T$, i.e., the first condition in Lemma~\ref{pro_Conditions} should be satisfied.
The maximum achievable revenue of the MBS manager with fixed and feasible quality assignment $\{w_t\}$ is given by
\begin{equation}\label{eqn_RevenueWithOptimalPricing}
R^*\big(\{w_t\}\big)=\max\limits_{\{p_t\}} \Bigg[ \sum\limits_{t=1}^T \Big(N_t \cdot p_t \Big) -  C\Big(\sum\limits_{t=1}^T N_t \cdot w_t\Big) \Bigg].
\end{equation}
From the above equation we can see that, the key point is to maximize $\sum_{t=1}^T \big(N_t \cdot p_t \big)$, since the cost function is constant with fixed quality assignment $\{w_t\}$.
Accordingly, we provide the following proposition for the optimal pricing strategy:

\vspace{-3mm}
\begin{proposition}\label{pro_OptimalPricing} \textbf{(Optimal Pricing Strategy) :}
Given that $\big\{ (w_t, p_t) \big\}$ is a feasible contract with feasible quality assignment $\{w_t\}$, the unique optimal pricing strategy $\{\hat{p}_t\}$ is:
\begin{equation}\label{eqn_OptimalPricing}
\left\{
\begin{array}{ll}
\hat{p}_1 = U(\lambda_1,w_1), &\\
\hat{p}_k = \hat{p}_{k-1} + U(\lambda_k,w_k) - U(\lambda_k,w_{k-1}), & \quad \forall k=2,3,\cdots,T.
\end{array}
\right.
\end{equation}
\end{proposition}

Its proof is given in Appendix~\ref{sec_Appendix_Pro4}.
According to this proposition, we write the general formula of the optimal prices $\{\hat{p}_t\}$ as
\vspace{-1mm}
\begin{equation}\label{eqn_OptimalPricingTerm}
\vspace{-1mm}
\hat{p}_t = U(\lambda_1,w_1)+\sum\limits_{i=1}^t \theta_i, \quad\quad \forall t=2,\cdots T,
\end{equation}
where $\theta_1=1$ and $\theta_i=U(\lambda_i,w_i)-U(\lambda_i,w_{i-1})$ for $i=2,\cdots T$.
The optimal pricing strategy is able to maximize $R$ and achieve $R^*$ with any given feasible quality assignment.
However, what $\{w_t\}$ is able to maximize $R^*$ and achieve the overall maximum value $\hat{R}$ is still unsolved.

\vspace{-4mm}
\subsection{Optimal Quality Assignment Problem}\label{sec_Design_Quality}

In this subsection, we analyze the optimal quality assignment problem based on the results in Section~\ref{sec_Design_Pricing}, and transform this problem into an easier form, as a preparation for the dynamic programming algorithm in Section~\ref{sec_Design_Algorithm}.

The optimal quality assignment problem is given by
\begin{equation}\label{eqn_OptimalQuality0}
\begin{array}{l}
\hat{R}  = \max\limits_{\{w_t\}} \Big[ R^*\big(\{w_t\}\big) \Big], \\
s.t. \sum\limits_{t=1}^T N_t w_t \le M,  \,\, w_1 \le w_2 \le \cdots \le w_T, \,\, \textrm{and } w_t=0,1,2\cdots
\end{array}
\end{equation}
where $R^*(\{w_t\})$ is the best revenue of a given quality assignment as given in~(\ref{eqn_RevenueWithOptimalPricing}).
Due to the optimal pricing $\{\hat{p}_t\}$ in~(\ref{eqn_OptimalPricingTerm}), we derive the expression of $R^*\big(\{w_t\}\big)$ as follows:
\begin{equation}\label{eqn_Revenue2}
\begin{array}{lll}
R^*\big(\{w_t\}\big) & = & \sum\limits_{t=1}^T \Big(N_t \cdot \hat{p}_t \Big) -  C\Big(\sum\limits_{t=1}^T N_t \cdot w_t\Big) = \sum\limits_{t=1}^T \bigg\{N_t \Big[U(\lambda_1,w_1)+\sum\limits_{i=1}^t \theta_i \Big] \bigg\} -  C\Big(\sum\limits_{t=1}^T N_t \cdot w_t\Big) \\
 & = & \sum\limits_{i=1}^T U(\lambda_i,w_i) \cdot \sum\limits_{t=i}^T N_t +\sum\limits_{i=1}^{T-1}\Big[ U(\lambda_{i+1},w_i) \cdot \sum\limits_{t=i+1}^T N_t \Big]  -  C\Big(\sum\limits_{t=1}^T N_t \cdot w_t\Big) \\
 & = & \sum\limits_{t=1}^{T} \bigg[ C_t \cdot U(\lambda_t,w_t)  - D_t \cdot U(\lambda_{t+1},w_t)  \bigg] -  C\Big(\sum\limits_{t=1}^T N_t \cdot w_t\Big), \\
\end{array}
\end{equation}
where $C_t = \Big(\sum\limits_{i=t}^T N_i \Big) $, $D_t = \Big( \sum\limits_{i=t+1}^T N_i \Big)$ for $t<T$, and $D_T = 0$.
Here, we are able to guarantee that $C_t>D_t \ge 0$, $\forall t=1,2,\cdots,T$, since $N_t>0, \forall t=1,2,\cdots,T$.
As we can observed from~(\ref{eqn_Revenue2}), $w_i$ and $w_j$ ($i \neq j$) are separated from each other in the first term.
This is a non-negligible improvement to find the best $\{w_t\}$.

\vspace{-3mm}
\begin{definition}\label{pro_Definition}
A set of functions $\Big\{G_t(w_t) \Big| \,  t=1, 2,\cdots T \Big\}$, with the quality $w_t$ as the independent variable of $G_t(\cdot)$, with $C_t$ and $D_t$ ($C_t>D_t \ge 0$) as the constants of $G_t(\cdot)$, is given by:
\begin{equation}\label{eqn_Gi}
G_t(w_t) =C_t \cdot U(\lambda_t,w_t) - D_t \cdot U(\lambda_{t+1},w_t), \quad w_t=0,1,2,\cdots \quad \forall t=1, 2,\cdots T.
\end{equation}
\end{definition}

Based on~(\ref{eqn_Revenue2}) and Definition~\ref{pro_Definition}, we have $R^*\big(\{w_t\}\big)=\sum\limits_{t=1}^{T}  G_t(w_t)   -  C\big(\sum\limits_{t=1}^T N_t \cdot w_t\big)$.
The meaning of $G_t(w_t)$ is the independent gain of setting $w_t$ for the $\lambda_t$-type UAVs regardless of the cost.

Based on $\{G_t(w_t)\}$, we can rewrite the optimization problem in~(\ref{eqn_OptimalQuality0}) as:
\begin{equation}\label{eqn_OptimalQuality1}
\begin{array}{l}
\hat{R}=\max\limits_{\{w_t\}} \Bigg[ \sum\limits_{t=1}^T G_t(w_t) - C\bigg( \sum\limits_{t=1}^T N_t w_t \bigg)  \Bigg] \\
s.t. \sum\limits_{t=1}^T N_t w_t \le M,  \,\, w_1 \le w_2 \le \cdots \le w_T, \,\, \textrm{and } w_t=0,1,2\cdots
\end{array}
\end{equation}
This problem can be further transformed into an equivalent one, given by
\begin{equation}\label{eqn_OptimalQuality2}
\begin{array}{l}
\hat{R}=\max\limits_{\{W=0,1,\cdots, M\}} \Bigg\{ \max\limits_{\{w_t\}} \bigg[ \sum\limits_{t=1}^T G_t(w_t) \bigg] - C\big( W \big) \Bigg\}, \\
s.t. \sum\limits_{t=1}^T N_t w_t \le W,  \,\, w_1 \le w_2 \le \cdots \le w_T, \,\, \textrm{and } w_t=0,1,2\cdots
\end{array}
\end{equation}
From this formulation, we can see that the optimal revenue can be acquired by trying each possible $W$ as the number of channels being sold.
For each fixed $W$, we only have to focus on how to maximize $\sum\limits_{t=1}^T G_t(w_t)$, given as
\begin{equation}\label{eqn_OptimalQuality3}
\begin{array}{l}
\max\limits_{\{w_t\}}  \sum\limits_{t=1}^T G_t(w_t), \\
s.t. \sum\limits_{t=1}^T N_t w_t \le W,  \,\, w_1 \le w_2 \le \cdots \le w_T, \,\, \textrm{and } w_t=0,1,2\cdots
\end{array}
\end{equation}
By calculating all the best results of~(\ref{eqn_OptimalQuality3}) with different possible values of $W$, we are able to obtain the optimal solution of~(\ref{eqn_OptimalQuality2}) by further taking into account $C(W)$.

Therefore, we regard~(\ref{eqn_OptimalQuality3}) as the key problem to be solved.
The proposed dynamic programming algorithm for this problem is presented in the next subsection.

\vspace{-4mm}
\subsection{Algorithm for the MBS Optimal Contract}\label{sec_Design_Algorithm}

In what follows, we first show the way of considering~(\ref{eqn_OptimalQuality3}) as a distinctive form of the knapsack problem~\cite{bib_Knapsack}, then provide our recurrence formula to calculate its maximum value $G_{max}$, next present the method to find the parameters $\{w_t\}$ that achieve $G_{max}$, and finally provide an overview of whole solution including the optimal quality assignment $\{\hat{w}_t\}$ and the optimal pricing $\{\hat{p}_t\}$.

\subsubsection{A special knapsack problem}\emph{}

First, we have to take a look at the constraints about the optimization parameters $\{w_i\}$.
Since $w_i=0,1,2\cdots$ and $\sum_{t=1}^T N_t w_t \!\le\! W$, we have $ w_t \!\le\! W$.
To distinguish from the notation of \emph{weight} in the following discussions, we use $K$ instead of $W$ as the common upper bound of $w_t$, $\forall t\!\in\![1,T]$, where $K\!\le\! W$.
And we rewrite the constraint as $ w_t \!\le\! K$.
Therefore for each~$t$, there are totally $K\!+\!1$ optional values of $w_t$, given by $\{0, 1 \cdots K\}$.
And the corresponding results of $G_t(w_t)$ are $\{G_t(0), G_t(1), G_t(2),\cdots , G_t(K)\}$, which represent the values of different object that we can choose.
In addition, we interpret the constraint $\sum_{t=1}^T N_t w_t \le W$ as the weight constraint in the knapsack problem, where $W$ is the weight capacity of the bag and setting $w_t=k$ means taking up the weight of $kN_t$.

\begin{table}[!thp]
\renewcommand\arraystretch{0.8}
\caption{All the optional objects to be selected}\label{tab_Algorithm} \centering
\vspace{-5mm}
\begin{tabular}{|p{5mm}|p{15mm}|p{55mm}|p{55mm}|}
\hline
 & Type & Optional Values & Corresponding Weights \\
\hline
1 & Type $\lambda_1$ & $G_1(0) \quad G_1(1) \quad G_1(2) \quad \cdots \quad G_1(K)$ & $0 \quad N_1 \quad 2N_1 \quad \cdots \quad KN_1$ \\
\hline
2 & Type $\lambda_2$ & $G_1(0) \quad G_2(1) \quad G_2(2)  \quad \cdots \quad G_2(K)$ & $0 \quad N_2 \quad 2N_2 \quad \cdots \quad KN_2$ \\
\hline
$\cdots$ & $\cdots$ & $\cdots$ & $\cdots$ \\
\hline
T & Type $\lambda_T$ & $G_T(0) \quad G_T(1) \quad G_T(2)  \quad \cdots \quad G_T(K)$ & $0 \quad N_T \quad 2N_T \quad \cdots \quad KN_T$ \\
\hline
\end{tabular}
\vspace{-2mm}
\end{table}

For the convenience of understanding, we list the values and the weight of different options in Table~\ref{tab_Algorithm}.
Each row presents all the options of a type and we should choose an option for each type.
And the $k^{th}$ option in the $t^{th}$ row provides us with the value of $G_t(k)$ and the weight of $kN_t$.
Due to the constraint of $w_1 \!\le\! w_2 \!\le\! \cdots \!\le\! w_T$, we cannot choose the $(k\!+\!1)^{th}, (k\!+\!2)^{th} \cdots$ options in the $t^{th}$ row if we have already chosen the $k^{th}$ option in the $(t\!+\!1)^{th}$ row.
Therefore, the algorithm introduced below is basically to start from the last row and end at the first row.

\subsubsection{The recurrence formula to calculate the maximum value $G_{max}$}\emph{}

The key nature of designing a dynamic programming algorithm is to find the sub-problems of the overall problem and write the correct recurrence formula.
Here we define $OPT(t,k,w)$, $\forall t\in[1,T]$, $\forall k\in[0,K]$ and $\forall w\in[0,W]$, as the optimal outcome that includes the decisions from the $T^{th}$ row to the $t^{th}$ row, with the conditions that 1) the $k^{th}$ option in the $t^{th}$ row is chosen and 2) the occupied weight is no more than $w$.
Since the algorithm starts from the $T^{th}$ row, we first provide the calculation of $OPT(T,k,w)$, $\forall k\in[0,K]$ and $\forall w\in[0,W]$, given as
\begin{equation}\label{eqn_Algorithm1}
OPT\big(T,k,w\big)=
\left\{
\begin{array}{ll}
G_T(k), & \textrm{if } w\ge k N_t, \\
-\infty, & \textrm{if } w< k N_t,
\end{array}
\right.
\end{equation}
where $-\infty$ implies that $OPT(T,k,w)$ is impossible to be achieved due to the lack of weight capacity.
This expression is straight forward since it only includes the $T^{th}$ row in Table~\ref{tab_Algorithm}.
From $OPT\big(T,k,w\big)$, we can calculate $OPT\big(t,k,w\big)$ for all $t\in[1,T-1]$, $k\in[0,K]$ and $w\in[0,W]$ by the following recurrence formula:
\begin{equation}\label{eqn_Algorithm2}
OPT\big(t,k,w\big)=
\left\{
\begin{array}{ll}
\max\limits_{l=k,\cdots,K} \Big[ G_t(k) +OPT\big(t+1,l,w-k N_t\big)  \Big], & \textrm{if } w\ge k N_t, \\
-\infty, & \textrm{if } w< k N_t.
\end{array}
\right.
\end{equation}
The meaning of this formula is:
If we want to choose $k$ in the $t^{th}$ row, then the option that made in the $(t+1)^{th}$ row must be within $[k,K]$ due to the constraint of $w_1 \le \cdots \le w_T$.
In addition, choosing $k$ in the $t^{th}$ row with total weight limit of $w$ indicates that there is only $w-k N_t$ left for the other rows from $t+1$ to $T$.
And if $w-k N_t < 0$, the outcome is $-\infty$ since choosing $k$ in the $t^{th}$ row is impossible.

Let $G_{max}$ denote $\max\limits_{\{w_t\}}  \sum\limits_{t=1}^T G_t(w_t)$, then we have the following expression:
\begin{equation}\label{eqn_Algorithm3}
G_{max} = \max\limits_{k=0\cdots K} \Big[ OPT\big(1,k,W\big) \Big].
\end{equation}
It means that we have to calculate $OPT(1,k,W)$ for all $k\in[0,K]$, which can be obtained by iteratively using~(\ref{eqn_Algorithm2}).

\subsubsection{The method to find the parameters $\{w_t\}$ that achieve $G_{max}$}\emph{}

Note that the above calculation only consider the value of the optimal result $G_{max}$.
To record what exact values of $\{w_t\}$ are chosen for this optimal result by the algorithm, we have to add another data structure, given as $D(t,k,w)$.
We let $D(t,k,w)=l$ if $OPT(t,k,w)$ chooses $l$ to maximize its value in the upper line of~(\ref{eqn_Algorithm2}), which is given by
\begin{equation}\label{eqn_Algorithm4}
D\big(t,k,w\big)=
\left\{
\begin{array}{ll}
arg\max\limits_{l=k,\cdots,K} \Big[ G_t(k) +OPT\big(t+1,l,w-k N_t\big)  \Big], & \textrm{if } w\ge k N_t, \\
0, & \textrm{if } w< k N_t.
\end{array}
\right.
\end{equation}
After acquiring $G_{max}$ in (\ref{eqn_Algorithm3}), we can use $D(t,k,w)$ to inversely find the optimal values of $\{w_t\}$ along the ``path" of the optimal solution.
Specifically, we have
\begin{equation}\label{eqn_Algorithm5}
\left\{
\begin{array}{lll}
\hat{w}_1 & = & arg \max\limits_{k=0\cdots K} \Big[ OPT\big(1,k,W\big) \Big], \\
\hat{w}_t & = & D\big( t-1,\hat{w}_{t-1},W-\sum\limits_{i=1}^{t-2} \hat{w}_i N_i\big), \quad \forall t=2, \cdots, T,
\end{array}
\right.
\end{equation}
where we define $\sum_{i=1}^{t-2} \hat{w}_i N_i $ as $0$ if $t\!-\!2 =0$, just for writing simplicity.

\subsubsection{An overview of whole solution}\emph{}

By now, we have presented the key part of our solution, i.e., the dynamic programming algorithm to solve the optimization problem in~(\ref{eqn_OptimalQuality3}).
The problem in~(\ref{eqn_OptimalQuality2}), i.e., our final goal, can be directly solved by setting different values of $W$ in~(\ref{eqn_OptimalQuality3}) and compare the corresponding results with the consideration of $C(W)$.

An overview of our entire solution is given in Algorithm~\ref{alg_Entire}.
It can be observed that the computational complexity of calculating $OPT(t,k,w)$ for all $k\!\in\![0,K]$, $w\!\in\![0,W]$ and $t\!\in\![1,T]$ is $O(TK^2W)$.
Therefore, the overall complexity is $O(MTK^2W)$, which can also be written as $O(TM^4)$ since $W\le M$ and $K \le W$.
Although $M^4$ seems to be non-negligible, there are usually no more than hundreds of available channels of a MBS to be allocated in practice.

\begin{algorithm}[!thp]
\caption{The algorithm of optimal contract for UAV offloading.}\label{alg_Entire}
\KwIn{Type information $\{\lambda_1, \cdots \lambda_T\}$, $\{N_1, \cdots N_T\}$, and the number of total channels $M$.}\vspace{-2mm}
\KwOut{Optimal pricing strategy $\{\hat{p}_1,\cdots \hat{p}_T\}$, optimal quality assignment $\{\hat{w}_1, \cdots \hat{w}_T\}$.}\vspace{-2mm}
\Begin
{\vspace{-2mm}
Calculate $G_t(k)$ for all $t\in[1,T]$ and $k\in[0,M]$ by~(\ref{eqn_Gi})\;\vspace{-2mm}
Calculate $C(m)$ for all $m\in[0,M]$ by~(\ref{eqn_Cost})\;\vspace{-2mm}
Initialize $\hat{R}=0$, $w_t=0$ for all $t\in[1,T]$, and $p_t=0$ for all $t\in[1,T]$\;\vspace{-2mm}
\For{$W$ is from $0$ to $M$}
     {\vspace{-2mm}
     Let $K=W$, to be the upper bound for each $w_i$\;\vspace{-2mm}
     Calculate $OPT(T,k,w)$ for $\forall k\in\![0,K]$ and $\forall w\in\![0,W]$ by~(\ref{eqn_Algorithm1})\;\vspace{-2mm}
     Calculate $OPT(t,k,w)$ for $\forall k\in\![0,K]$, $\forall w\in\![0,W]$ and $\forall t\in\![1,T\!-\!1]$ by~(\ref{eqn_Algorithm2})\;\vspace{-2mm}
     Acquire $G_{max}$ from $\{OPT(1,w,t)\}$ according to~(\ref{eqn_Algorithm3})\;\vspace{-2mm}
     \If{$G_{max}-C(W)>\hat{R}$}
     {\vspace{-2mm}
     Update the overall maximum revenue $\hat{R}=G_{max}-C(W)$\;\vspace{-2mm}
     Update $\hat{w}_t$ for all $t\in[1,T]$ according to~(\ref{eqn_Algorithm4}) and~(\ref{eqn_Algorithm5})\;\vspace{-2mm}
     Update $\hat{p}_t$ for all $t\in[1,T]$ based on $\{\hat{w}_t\}$ according to~(\ref{eqn_OptimalPricing})\;\vspace{-2mm}
     }\vspace{-2mm}
     }\vspace{-2mm}
}\vspace{-2mm}
\end{algorithm}

\vspace{-4mm}
\subsection{Social Optimal Contract}\label{sec_Design_Social}

\begin{figure}[!thp]
\centering
\includegraphics[width=4.0in]{{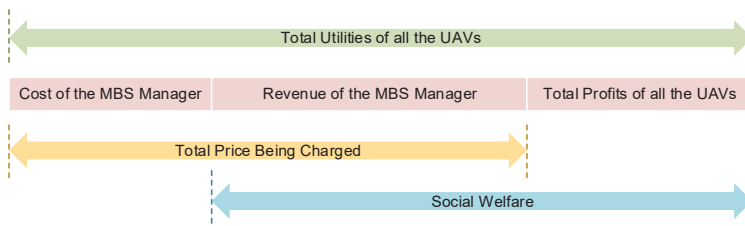}}
\vspace{-5mm}
\caption{The relation of the social welfare, the revenue of the MBS manager, and the total profit of the UAVs operators.}\label{fig_SocialWelfare}
\vspace{-5mm}
\end{figure}

To better discuss the effectiveness of the above MBS optimal contract, in the following, we briefly discuss another contract that aims to maximize social welfare.
In our context, social welfare indicates the sum of the revenue of the MBS and the total profits of the UAVs, which also means the increase of the number of users that can be served by the overall system.
The objective is given by
\begin{equation}\label{eqn_SocialObjective}
\hat{S} = \max\limits_{\{w(\lambda)\} , \{p(\lambda)\}} \sum\limits_{\lambda \in \Lambda} \Big(N_\lambda \cdot U\big(\lambda, w(\lambda)\big) \Big) -  C\Big(\sum\limits_{\lambda \in \Lambda} N_\lambda \cdot w(\lambda)\Big),
\end{equation}
where the first term is the total utility of the UAVs, the second term is the cost of the MBS, and we omit the constraints since they are the same with those in~(\ref{eqn_ObjectiveFunction}).
This optimization problem has a similar structure with~(\ref{eqn_ObjectiveFunction}) and can be solved by the proposed dynamic programming algorithm with only minor changes.
To calculate the optimal $\{ w(\lambda) \}$ and $\{ p(\lambda) \}$, we need to replace $G_t(k)$ by $N_t U(t,k)$ in line 2 of Table~\ref{tab_Algorithm}.
In addition, we use $U_{max}$ to replace $G_{max}$ to represent the maximum overall utility of the UAVs.
At last, the equation in line 11 of Table~\ref{tab_Algorithm} should be replaced by $\hat{S}=U_{max}\!-\!C(W)$ to represent the maximum social welfare.
For writing convenience, in the rest part of this paper, we call the solutions of~(\ref{eqn_ObjectiveFunction}) and~(\ref{eqn_SocialObjective}) as the ``MBS optimal contract" and the ``social optimal contract", respectively.
In addition, the relation of social welfare and MBS's revenue is illustrated in Fig.~\ref{fig_SocialWelfare}.

\section{Theoretical Analysis and Discussions}\label{sec_DiscussionHeight}

In this section, we briefly discuss the impact of the height of the UAVs, $H$.
Since $H$ influences the optimal revenue of the MBS $\hat{R}$, through the types of the UAVs $\{\lambda_t\}$, we first discuss the impact of $H$ on $\{\lambda_t\}$ in Section~\ref{sec_DiscussionHeight_1} and then discuss the impact of $\{\lambda_t\}$ on $\hat{R}$ in Section~\ref{sec_DiscussionHeight_2}.

\vspace{-4mm}
\subsection{The Impact of the Height on the UAV Types}\label{sec_DiscussionHeight_1}

\begin{proposition}\label{pro_OptimalHeight}
With fixed transmission power $P_{UAV}$ and $P_{MBS}$, fixed terrain parameters $a$, $b$, $\eta_{Los}$ and $\eta_{NLoS}$, fixed average active user density $\sigma_n$, fixed horizontal locations of the UAVs, and unified height $H\in [0,+\infty)$ of the UAVs, there exists a height $\hat{H}_n$ that can maximize the effective offloading area of UAV$_n$.
\end{proposition}
\vspace{-3mm}

The proof of this proposition is given in Appendix~\ref{sec_Appendix_Pro5}.
From this proposition, we know that in the process of $H$ varying from $0$ to $+\infty$, different UAVs are able to achieve their maximum effective offloading areas at different heights.
However, if all the UAVs are horizontally symmetrically distributed around the MBS (as shown in Fig.~\ref{fig_Height} in Section~\ref{sec_Simulation_Results}), their optimal heights will be the same since the UAVs have symmetrical positions.
Therefore, there is a globally optimal height $\hat{H}$ that can maximize $S_n$, for all $n\in\{1,2,\cdots N\}$.
Due to the fact that the types of the UAVs is given by $\lambda_n=\sigma_n S_n$, we can also achieve the largest type for each UAV.

\vspace{-4mm}
\subsection{The Impact of the UAV Types on the Optimal Revenue}\label{sec_DiscussionHeight_2}

For any two random sets of types $\{\lambda_1,\cdots \lambda_{T_1}\}$ and $\{\lambda_1^\prime,\cdots \lambda_{T_2}^\prime\}$, there is no obvious relation of the outcomes of the corresponding two MBS optimal contracts.
However, some properties can be explored when we add some constraints, as given in the following proposition:

\vspace{-3mm}
\begin{proposition}\label{pro_OptimalTypes}
Given a fixed number of types $T$, two sets of types $\{\lambda_t\}$, $\{\lambda_t^{\prime}\}$, and the constraint $\lambda_t\le \lambda_t^{\prime}$, $\forall t \in[1,T]$, we have $\hat{R}\le \hat{R}^{\prime}$, where $\hat{R}$ is the MBS's revenue of a MBS optimal contract with inputs $\{\lambda_t\}$ and $\hat{R}^{\prime}$ is the MBS's revenue of a MBS optimal contract with inputs $\{\lambda_t^{\prime}\}$.
\end{proposition}
\vspace{-3mm}

The proof of this proposition is given in Appendix~\ref{sec_Appendix_Pro6}.
With Proposition~\ref{pro_OptimalHeight} and Proposition~\ref{pro_OptimalTypes}, we can directly obtain a conclusion that, there exists a highest value of the MBS's revenue by manipulating the height of the UAVs, as long as the UAVs are horizontally symmetrically distributed around the MBS, as shown in Section~\ref{sec_Simulation_Results}.

\section{Simulation Results}\label{sec_Simulation}

In this section, we simulate and compare the outcomes of the MBS optimal contract and the social optimal contract under different settings.
Simulation setups are given in Section~\ref{sec_Simulation_Setups}, simulation results and corresponding discussions are provided in Section~\ref{sec_Simulation_Results}.

\vspace{-4mm}
\subsection{Simulation Setups}\label{sec_Simulation_Setups}

We set $M$ within $[100, 300]$, which is sufficient to generally evaluate a real system such as LTE~\cite{bib_LTE}.
The terrain parameters are set as $a=\!11.95$ and $b=\!0.136$, indicating a typical urban environment.
We also set the transmission power as $P_{UAV}\!<\!P_{MBS}$, due to the typical consideration of UAVs that they have limited battery capacities.
Details of the settings of all the parameters can be found in Table~\ref{tab_Parameters}.

\begin{table}[!thp]
\renewcommand\arraystretch{0.8}
\caption{Simulation parameters}\label{tab_Parameters} \centering
\vspace{-5mm}
\begin{tabular}{|p{75mm}|p{40mm}|}
\hline
Terrain parameters $a$ and $b$ & $11.95$ and $0.136$ \\
\hline
Additional pathloss parameters $\eta_{NOLS}$ and $\eta_{LOS}$ & $2$ and $20$ \\
\hline
Transmission power $P_{MBS}$ and $P_{UAV}$ & $10W$ and $50mW$ \\
\hline
Downlink transmission frequency $f$ & $3 GHz$ \\
\hline
Height of UAVs $H$ ($m$) & between $200$ and $1000$ \\
\hline
Average active user density $\sigma_n$ ($km^{-2}$) & between $10$ and $20$ \\
\hline
Number of UAVs' types $T$ & between $1$ and $20$ \\
\hline
Number of each type of UAVs $\{N_t\}$ & between $1$ and $10$ \\
\hline
Average active user number of UAVs $\{\lambda_t\}$ & between $1$ and $20$ \\
\hline
Average active user number of MBS $\lambda_{BS}$ & between $10$ and $200$ \\
\hline
Number of total channels of MBS $M$ & between $100$ and $300$ \\
\hline
\end{tabular}
\vspace{-5mm}
\end{table}

In the following simulations, we first study the UAV offloading system based on given UAV types (i.e., fixed active user number for each UAV), from which we can acquire basic comprehension of the MSB optimal contract and the social optimal contract.
Then we further study a more practical scenario where the height of the UAVs determines the types of them.

\vspace{-4mm}
\subsection{Simulation Results and Discussions}\label{sec_Simulation_Results}

\begin{figure}[!thp]
\centering
\includegraphics[width=5in]{{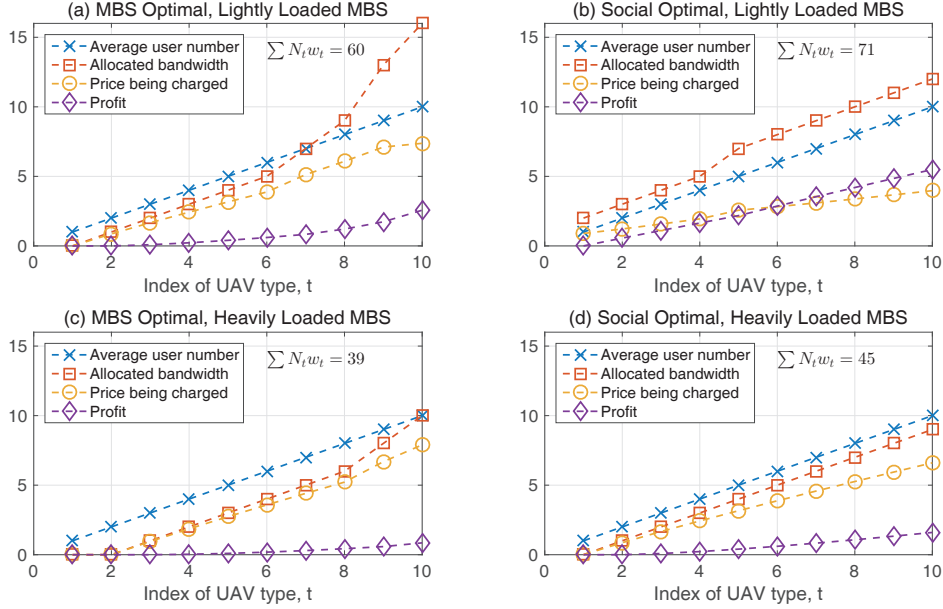}}
\vspace{-5mm}
\caption{The structure of the optimal contracts where $T=10$, $\{N_t\}=(1,1,\cdots 1)$, $\{\lambda_t\}=(1,2,\cdots 10)$, and $M=200$, with $\lambda_{BS}=120$ for (a) and (b), and $\lambda_{BS}=160$ for (c) and (d). In addition, (a) and (c) show MBS optimal contracts while (b) and (d) show social optimal contracts.}\label{fig_Basic}
\vspace{-5mm}
\end{figure}

We first illustrate the typical structure of the contract that designed according to our algorithm, as given in Fig.~\ref{fig_Basic}, where we set $T=10$, $\{N_t\}=(1,1,\cdots 1)$, $\{\lambda_t\}=(1,2,\cdots 10)$, and $M=200$.
All the four subplots show the patterns of $\{w_t\}$, $\{p_t\}$, and $\{U(t,w_t)-p_t\}$ with respect to different type $\lambda_t$.
To be specific, subplots (a) and (b) show the results of lightly loaded MBS ($\lambda_{BS}=120$) while (c) and (d) show the results of heavily loaded MBS ($\lambda_{BS}=160$).
In addition, subplots (a) and (c) are the outcomes of MBS optimal contracts while (b) and (d) are the outcomes of social optimal contracts.
In any one of these subplots, we can see that a higher type of UAV is allocated with more channels but also a higher price.
It can also be observed that a higher type gains more profit compared with a lower type, i.e., $U(i,w_i)\!-\!p_i \le U(j,w_j)\!-\!p_j$ as long as $i < j$.
In Fig.~\ref{fig_Basic} (a),  it is noticeable that for $\lambda_8$, $\lambda_9$ and $\lambda_{10}$-types, the allocated channels exceed their respective average user numbers.
Such phenomenon is quite reasonable since a UAV needs more channels $w$ than its average active user number $\lambda$ to deal with the situation of burst access.
And due to the IP property, higher types consider additional channels more valuable than lower types.
Therefore, only $\lambda_8$, $\lambda_9$ and $\lambda_{10}$-types are allocated with excessive channels.
By comparing Fig.~\ref{fig_Basic} (a) with Fig.~\ref{fig_Basic} (b), or Fig.~\ref{fig_Basic} (c) with Fig.~\ref{fig_Basic} (d), we find that a social optimal contract allocates more channels than a MBS optimal contract, where we have $60$ against $71$ in (a) and (b), and $39$ against $45$ in (c) and (d).
It can be considered that a social optimal contract is more ``generous'' than a MBS optimal contract.
By comparing Fig.~\ref{fig_Basic} (a) with Fig.~\ref{fig_Basic} (c), or Fig.~\ref{fig_Basic} (b) with Fig.~\ref{fig_Basic} (d),  we can also find the difference of the numbers of totally allocated channels.
This is because the cost of a heavily loaded MBS allocating the same number of channels is greater than that of a lightly loaded MBS.

\begin{figure}[!thp]
\centering
\includegraphics[width=5in]{{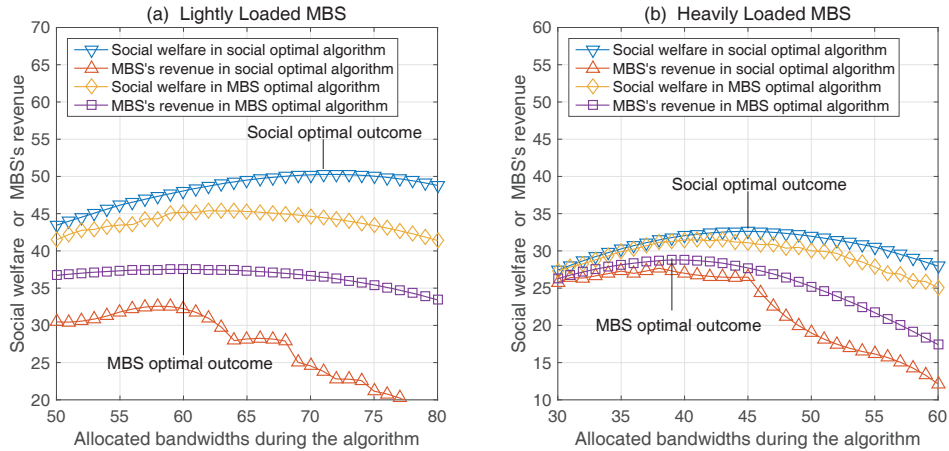}}
\vspace{-5mm}
\caption{The change of social welfare and MBS's revenue during the social optimal algorithm and MBS optimal algorithm, where $T=10$, $\{N_t\}=(1,1,\cdots 1)$, $\{\lambda_t\}=(1,2,\cdots 10)$, $M=200$, with $\lambda_{BS}=120$ for (a) and $\lambda_{BS}=160$ for (b).}\label{fig_Bandwidth}
\vspace{-5mm}
\end{figure}

To better explain the aforementioned bandwidth differences, we provide Fig.~\ref{fig_Bandwidth} to show how social welfare and the MBS's revenue change during the algorithm with $W$ setting from $0$ to $M$ (as described in line 5 in Table~\ref{tab_Algorithm}).
In Fig.~\ref{fig_Bandwidth} (a), the upmost blue curve shows the change of social welfare during the social optimal algorithm.
The highest point of this curve represents the corresponding social optimal contract, which makes $W=71$ just as given in Fig.~\ref{fig_Basic} (b).
The lowermost orange curve shows the corresponding change of the MBS's revenue during the social optimal algorithm.
For the MBS optimal algorithm, the resulting curve of the the MBS's revenue lies above the orange one from the social optimal algorithm, while the resulting curve of social welfare lies below the blue one from the social optimal algorithm.
Since the two groups of curves do not coincide, we can deduce that the structure of the solutions of the two algorithms are not identical.
For a fixed $W$, the MBS optimal algorithm somehow changes the allocation of channels among different types to increase the MBS's revenue, which results in a reduction of social welfare.
And the bandwidth allocation of the MBS optimal contract is $W=60$,  just as given in Fig.~\ref{fig_Basic} (a).
In Fig~\ref{fig_Bandwidth} (b), we also show the situation of heavily loaded MBS, where the relation of these curves are similar, as well as the reason that causes this.

\begin{figure}[!thp]
\centering
\includegraphics[width=5in]{{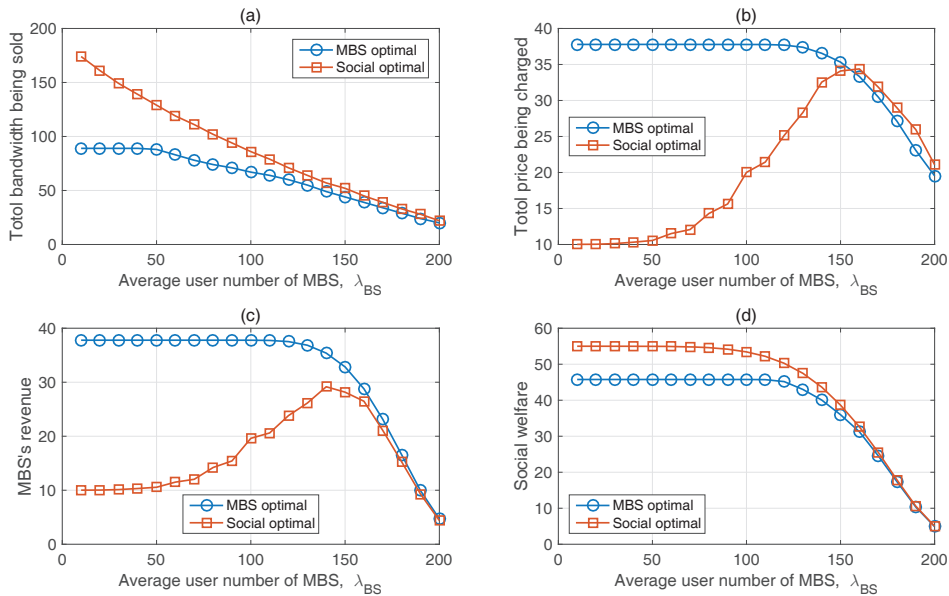}}
\vspace{-5mm}
\caption{The impacts of $\lambda_{BS}$, where $T=10$, $\{N_t\}=(1,1,\cdots 1)$, $\{\lambda_t\}=(1,2,\cdots 10)$ and $M=200$.}\label{fig_LambdaBS}
\vspace{-5mm}
\end{figure}

Fig.~\ref{fig_LambdaBS} illustrates the impacts of $\lambda_{BS}$ (i.e., the load of the MBS) on the different part of the utility of the whole system as presented in Fig.~\ref{fig_SocialWelfare}.
From Fig.~\ref{fig_LambdaBS} (a) we can see that, the difference of allocated channels between the MBS optimal contract and the social optimal contract becomes smaller as the load of MBS gets heavier.
This is due to the fact that the cost of MBS rises fast when it is heavily loaded and neither the MBS optimal or the social optimal contract can allocate enough channels as desired.
Fig.~\ref{fig_LambdaBS} (b) shows us that the MBS optimal contract is able to guarantee a high level of total prices that being charged as the MBS is not heavily loaded.
In addition, the total prices being charged according to the social optimal contract is not monotonous and may rapidly change.
For the case $\lambda_{BS}\!>\!150$, although the total price being charged in the MBS optimal contract is lower than that in the social optimal contract, the final revenue of the MBS is still higher in the MBS optimal contract as shown in Fig.~\ref{fig_LambdaBS} (c).
This is because the MBS optimal contract has less total bandwidth being sold, which reduces the cost of the MBS.
The social welfare is given in Fig.~\ref{fig_LambdaBS} (d), which implies that for both MBS and social optimal contracts, a heavier loaded MBS could bring a lower overall system efficiency.

\begin{figure}[!thp]
\centering
\includegraphics[width=6in]{{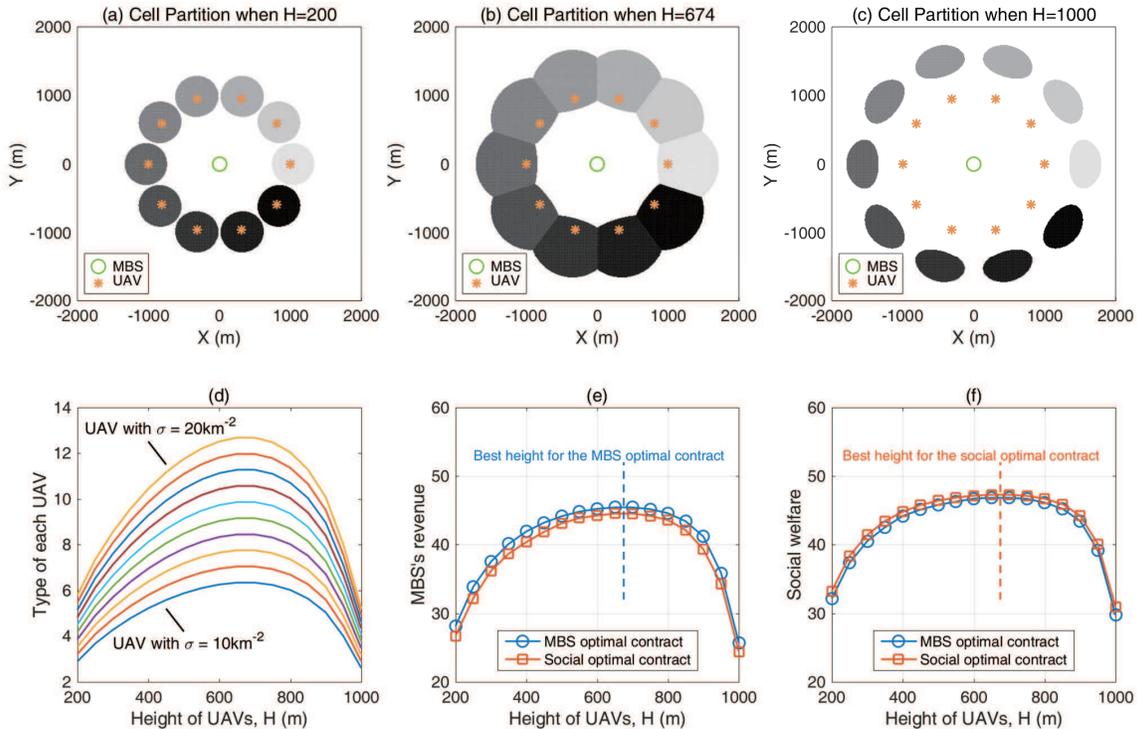}}
\vspace{-5mm}
\caption{The impact of the height of UAVs. The subplots (a), (b) and (c) show the top views of the cell partition with different height settings. The white areas represent MBS's effective service regions, while gray areas represent UAVs' effective offloading regions. The subplot (d) provides the impact on the type of each UAV with different active user density. The subplots (e) and (f) illustrate the impacts of the height of UAVs on ``MBS's revenue" and ``social welfare", respectively.}\label{fig_Height}
\vspace{-5mm}
\end{figure}

Finally, we take a look at the impact of the height of the UAVs, as presented in Fig.~\ref{fig_Height}, where $M=200$, $\lambda_{BS}=150$.
The considered $10$ UAVs are located $1000m$ horizontally from the MBS and symmetrically distributed.
The average active user density of the effective offloading region of UAV$_n$ (i.e., $\sigma_n$) is set from $10 km^{-2}$ to $20 km^{-2}$.
From the top three subplots in Fig.~\ref{fig_Height}, we can see that the offloading regions of these UAVs first expand then shrink when the height of the UAVs monotonously increases.
The maximum offloading areas can be achieved at $H\!=\!674$, where the UAVs can cover the largest number of active users, as given in Fig.~\ref{fig_Height} (d).
In addition, the MBS's revenue can be maximized when offloading areas become the largest, as discussed in Section~\ref{sec_DiscussionHeight}.
It can also be observed in Fig.~\ref{fig_Height} (f) that the profile of the social welfare in the MBS optimal contract is very close to that of that social welfare in the social optimal contract.
In addition, the best height for the social optimal contract ($H\!=\!676$) is very close to the best height for the MBS optimal contract ($H\!=\!674$).
Therefore, we can infer that, the height $H$ that designated by the selfish MBS manager will generally keep a high social welfare.
In other words, the performance of the overall system will not be significantly impaired.

\section{Conclusion}\label{sec_Conclusion}

In this paper, we focused on the scenario where the UAVs were deployed in a cellar network to better serve local mobile users.
Considering the selfish MBS manager and the selfish UAV operators, we modeled the utilities and the costs of spectrum trading among them and formulated the problem of designing the optimal contract for the MBS manager.
To deduce the optimal contract, we first derived the optimal pricing strategy based on a fixed quality assignment, and then analyze and transform the optimal quality assignment problem, in which way it can be solved by the proposed dynamic programming algorithm in polynomial time.
In the simulations, by comparing with the social optimal contract, we found that the MBS optimal contract allocated fewer channels to the UAVs to guarantee a lower level of costs.
Moreover, the best height of the UAVs for the selfish MBS manager can keep a high performance of the overall system.

\begin{appendices}\label{sec_Appendix}

\section{Proof of Lemma 1}\label{sec_Appendix_Lem1}

\begin{IEEEproof}
Consider $X_\alpha$ as a Poisson distribution random variable with mean value $\alpha$, we have $P(X_\alpha \ge k) = 1- P(X_\alpha < k) = 1 - e^{-\alpha} \sum\limits_{i=0}^{k-1} \dfrac{\alpha^i}{i!}$.
Since $\alpha$ can be a real number in its definition domain, we derive the derivative of $P(X_\alpha \ge k)$ with respect to $\alpha$, given as
\begin{equation*}
\dfrac{\partial P(X_\alpha \ge k)}{\partial \alpha} = e^{-\alpha} \sum\limits_{i=0}^{k-1} \dfrac{\alpha^i}{i!} - e^{-\alpha} \dfrac{\partial}{\partial \alpha}\bigg( \sum\limits_{i=0}^{k-1} \dfrac{\alpha^i}{i!} \bigg).
\end{equation*}
For $k=1$, $\dfrac{\partial}{\partial \alpha}\big( \sum\limits_{i=0}^{k-1} \dfrac{\alpha^i}{i!} \big)=0$. And for $k>1$, $\dfrac{\partial}{\partial \alpha}\big( \sum\limits_{i=0}^{k-1} \dfrac{\alpha^i}{i!} \big)=\sum\limits_{i=1}^{k-1} \dfrac{\alpha^{i-1}}{(i-1)!}=\sum\limits_{i=0}^{k-2} \dfrac{\alpha^{i}}{i!}$.
Therefore, we have $\dfrac{\partial P(X_\alpha \ge k)}{\partial \alpha} = e^{-\alpha} \dfrac{\alpha^{k-1}}{(k-1)!} >0$, $\forall k\in \mathbb{Z}^+$ and $\alpha >0$.
With any given $\lambda > \lambda^\prime > 0$, we can deduce that $P(X_{\lambda} \ge k) - P(X_{\lambda^\prime} \ge k) = \int_{\lambda^\prime}^{\lambda} \dfrac{\partial P(X_\alpha \ge k)}{\partial \alpha} d\alpha >0$, $\forall k\in \mathbb{Z}^+$.
\end{IEEEproof}

\section{Proof of Proposition 1}\label{sec_Appendix_Pro1}

\begin{IEEEproof}
Consider a fixed value $w \!\in\! \mathbb{N}$ and $\lambda\!>\!\lambda^\prime\!>\!0$.
If $w\!=\!0$, we have $U(\lambda,w)\!=\!U(\lambda^\prime,w)=0$ according to the definition.
If $w\!>\!0$, then $U(\lambda,w) - U(\lambda^\prime,w) \!= \!\sum\limits_{k=1}^{k=w} \big[ P(X_{\lambda} \ge w) - P(X_{\lambda^\prime} \ge w) \big] \!>\!0$ according to Lemma~\ref{pro_Poisson}.
Therefore, $U(\lambda,w)$ monotonously increases with $\lambda$.

Now consider a fixed $\lambda>0$ and $\forall w>w^\prime \ge 0$, where $w,w^\prime \in \mathbb{N}$.
We have $U(\lambda,w)-U(\lambda,w^\prime)=P(X_{\lambda} \ge w) +\cdots + P(X_{\lambda} \ge w^\prime+1) \ge P(X_{\lambda} \ge w) >0$.
Therefore, $U(\lambda,w)$ monotonously increases with $w$.

For a fixed $\lambda>0$ and $\forall w \ge 1$, we have $U^\prime(w)=U(\lambda,w)-U(\lambda,w-1)=P(X_{\lambda} \ge w)$.
And for $w \ge 2$, we have $U^{\prime\prime}(w)=U^\prime(w)-U^\prime(w-1)=-P(X_{\lambda} = w-1) <0$.
Therefore, the marginal increase of $U(\lambda,w)$ with respect to $w$ gets smaller as $w$ increases.
\end{IEEEproof}

\section{Proof of Proposition 2}\label{sec_Appendix_Pro2}

\begin{IEEEproof}
According to the definitions of the utility function in (\ref{eqn_UtilityDefinition1}) and (\ref{eqn_UtilityDefinition2}), we have
\begin{equation}\label{eqn_PropositionIP_1}
U(\lambda,w)-U(\lambda,w^\prime)=P(X_{\lambda} \ge w) +\cdots + P(X_{\lambda} \ge w^\prime+1),
\end{equation}
\begin{equation}\label{eqn_PropositionIP_2}
U(\lambda^\prime,w)-U(\lambda^\prime,w^\prime)=P(X_{\lambda^\prime} \ge w) +\cdots + P(X_{\lambda^\prime} \ge w^\prime+1).
\end{equation}
Based on Lemma~\ref{pro_Poisson}, each term in (\ref{eqn_PropositionIP_1}) is greater than each corresponding term in (\ref{eqn_PropositionIP_2}).
Therefore we can obtain $U(\lambda,w)-U(\lambda,w^\prime)>U(\lambda^\prime,w)-U(\lambda^\prime,w^\prime)$.
\end{IEEEproof}

\section{Proof of Lemma 2}\label{sec_Appendix_Lem2}

\begin{IEEEproof}
\textbf{Necessity:}
These $3$ conditions can be deduced from the IC \& IR constraints and the IP property as follows:
1) Since $\{\lambda_1, \lambda_2, \cdots \lambda_T\}$ is written in the ascending order, we have $0 \le w_1 \le w_2 \le \cdots \le w_T$ and $0 \le p_1 \le p_2 \le \cdots \le p_T$ according to Proposition~\ref{pro_PreviousWork}, where $w_i=w_{i+1}$ if and only if $p_i=i_{i+1}$.
2) Considering the IR constraint of $\lambda_1$-type UAVs, we can directly obtain $0 \le p_1 \le U(\lambda_1,w_1)$.
Here, if $w_x=0$, then $U(\lambda_t,w_t)=0$ and $p_t=0$ for any $t \le x$.
3) Considering the IC constraint for the $k$-type and the $(k-1)$-type where $k>1$, the corresponding expressions are given by $U(\lambda_k,w_k)-p_k \ge U(\lambda_k,w_{k-1})-p_{k-1}$, and $U(\lambda_{k-1},w_{k-1})-p_{k-1} \ge U(\lambda_{k-1},w_{k})-p_{k}$.
As we focus on the possible scope of $p_k$, we can deduce that $p_{k-1}+\big[U(\lambda_{k-1},w_{k})-U(\lambda_{k-1},w_{k-1})\big] \le p_k \le p_{k-1} +\big[U(\lambda_k,w_k)-U(\lambda_k,w_{k-1})\big]$.

\textbf{Sufficiency:} We have to prove that the prices $\{ (p_t) \}$ determined by these conditions satisfy the IC and IR constraints.
And the basic idea is to use mathematical induction, from $(w_1, p_1)$ to $(w_T, p_T)$, by adding the quality-price terms once at a time into the whole contract.
For writing simplicity, the contract that only contains the first $k$ types of UAVs is denoted as $\Psi(k)$, where $\Psi(k)=\big\{ (w_t, p_t) \big\}, 1\le t \le k$.
First, we can verify that $w_1 \ge 0$ and $0<p_i<U(\lambda_1,w_1)$ provided by the above conditions is feasible in $\Psi(1)$, since the IR constraint $U(\lambda_1,w_1) - p_i > 0$ is satisfied and the IC constraint is not useful in a single-type contract.

In the rest part of our proof, we show that if $\Psi(k)$ is feasible, then $\Psi(k+1)$ is also feasible, where $k+1 \le T$.
To this end, we need to prove that (1) the newly added $\lambda_{k+1}$-type complies with its IC and IR constraints, given by
\begin{equation}\label{eqn_Lamma_1}
\left\{
\begin{array}{lll}
U(\lambda_{k+1},w_{k+1})-p_{k+1} & \ge & U(\lambda_{k+1},w_i)-p_i, \quad \forall i=1,2,\cdots, k,\\
U(\lambda_{k+1},w_{k+1})-p_{k+1} & \ge & 0 ,
\end{array}
\right.
\end{equation}
and (2) the existing $k$ types still comply with their IC constraints with the addition of $\lambda_{k+1}$-type, given by
\begin{equation}\label{eqn_Lamma_2}
U(\lambda_i,w_i)-p_i \ge U(\lambda_{i},w_{k+1})-p_{k+1}, \quad \forall i=1,2,\cdots, k.
\end{equation}

\emph{First, we prove~(\ref{eqn_Lamma_1}):} Since $\Psi(k)$ is feasible, the IC constraint of $\lambda_k$-type should be satisfied, given by $U(\lambda_k,w_i)-p_i \le U(\lambda_k,w_k)-p_k, \forall i=1,2,\cdots, k$.
Based on the right inequality in the third condition, we have $p_{k+1} \le p_k + U(\lambda_{k+1},w_{k+1}) -U(\lambda_{k+1},w_k)$.
By adding up these two inequalities, we have $U(\lambda_k,w_i)-p_i + p_{k+1} \le U(\lambda_k,w_k) + U(\lambda_{k+1},w_{k+1}) - U(\lambda_{k+1},w_k), \forall i=1,2,\cdots, k$.
According to the IP property, we can obtain that $U(\lambda_k,w_k) - U(\lambda_k,w_i) \le U(\lambda_{k+1},w_k) - U(\lambda_{k+1}, w_i), \forall i=1,2,\cdots, k$, since $\lambda_{k+1} > \lambda_{k}$ and $ w_k \ge w_i$.
Again, by combining these two inequalities together, we can prove the IC constraint of the $\lambda_{k+1}$-type, given by $U(\lambda_{k+1},w_{k+1})-p_{k+1} \ge U(\lambda_{k+1},w_i)-p_i, \forall i=1,2,\cdots, k$.
The IR constraint of the $\lambda_{k+1}$-type can be easily deduced from the above IC constraint since $U(\lambda_{k+1},w_i)-p_i \ge U(\lambda_i,w_i)-p_i \ge 0, \forall i=1,2,\cdots, k$.
And therefore, we have $U(\lambda_{k+1},w_{k+1})-p_{k+1} \ge 0$.

\emph{Then, we prove~(\ref{eqn_Lamma_2}):} Since $\Psi(k)$ is feasible, the IC constraint of $\lambda_i$-type, $i=1,2,\cdots, k$, should be satisfied, given by $U(\lambda_i,w_k)-p_k \le U(\lambda_i,w_i)-p_i, \forall i=1,2,\cdots, k$.
Based on the left inequality in the third condition, we have $p_k + U(\lambda_k,w_{k+1})-U(\lambda_k,w_k) \le p_{k+1}$.
By adding up the above two inequalities, we have $U(\lambda_i,w_k) + U(\lambda_k,w_{k+1}) - U(\lambda_k,w_k) \le U(\lambda_i,w_i) - p_i +p_{k+1} \forall i=1,2,\cdots, k$.
According to the IP property, we can obtain that $U(\lambda_i,w_{k+1}) - U(\lambda_i, w_k) \le U(\lambda_k,w_{k+1}) - U(\lambda_k,w_k), \forall i=1,2,\cdots, k$,since $\lambda_k \ge \lambda_i$ and $ w_{k+1} \ge w_k$.
Again, by combining the above two inequalities together, we can prove the IC constraint of the existing types, $\lambda_i$, $\forall i=1,2,\cdots, k$, given by $U(\lambda_i,w_i)-p_i \ge U(\lambda_{i},w_{k+1})-p_{k+1}$.

So far, we have proved that $\Psi(1)$ is feasible, and if $\Psi(k)$ is feasible then $\Psi(k+1)$ is also feasible.
We can conclude that the final contract $\Psi(T)$ which includes all the types is feasible.
Therefore, these three necessary conditions are also sufficient conditions.
\end{IEEEproof}

\section{Proof of Proposition 4}\label{sec_Appendix_Pro4}

\begin{IEEEproof}
By comparing~(\ref{eqn_OptimalPricing}) with Lemma~\ref{pro_Conditions}, we can find that $\{\hat{p}_t\}$ is a feasible pricing strategy.
In the following, we first prove that $\{\hat{p}_t\}$ is optimal, then prove that it is unique.

\emph{Optimality:}
In the condition that quality assignment $\{w_t\}$ is fixed, $\{\hat{p}_t\}$ is optimal if and only if $\sum_{t=1}^T \big(N_t \cdot \hat{p}_t \big) \ge \sum_{t=1}^T \big(N_t \cdot p_t \big)$, where $\{p_t\}$ is any pricing strategy that satisfies the conditions in Lemma~\ref{pro_Conditions}.
Let's assume that there exists another better strategy $\{\tilde{p}_t\}$ for the MBS manager, i.e., $\sum_{t=1}^T \big(N_t \cdot \tilde{p}_t \big) \ge \sum_{t=1}^T \big(N_t \cdot \hat{p}_t \big)$.
Since $N_t>0$ for all $t=1,2,\cdots,T$, there is at least one $k \in \{1,2,\cdots,T\}$ that satisfies $\tilde{p}_k > \hat{p}_k$.
To guarantee that $\{\tilde{p}_t\}$ is still feasible, the following inequality must be complied with according to Lemma~\ref{pro_Conditions}: $\tilde{p}_k \le \tilde{p}_{k-1} +U(\lambda_k,w_k) - U(\lambda_k,w_{k-1}),\textrm{if } k>1$.
Since $\tilde{p}_k > \hat{p}_k$, we have $\hat{p}_k < \tilde{p}_{k-1} +U(\lambda_k,w_k) - U(\lambda_k,w_{k-1}),  \textrm{if } k>1$.
By substituting~(\ref{eqn_OptimalPricing}) into the above inequality, we have $\tilde{p}_{k-1} > \hat{p}_k + U(\lambda_k,w_k) - U(\lambda_k,w_{k-1}) = \hat{p}_{k-1}, \textrm{if } k>1$.
Repeat this process and we can finally obtain the result that $\tilde{p}_1>\hat{p}_1=U(\lambda_1,w_1)$, which contradicts with Lemma~\ref{pro_Conditions} where $p_1$ should not exceed $U(\lambda_1,w_1)$.
Due to this contradiction, the above assumption that $\{\tilde{p}_t\}$ is better than $\{\hat{p}_t\}$ is impossible.
Therefore, $\{\hat{p}_t\}$ is the optimal pricing strategy for the MBS manager.

\emph{Uniqueness:}
Assume that there exists another pricing strategy $\{\tilde{p}_t\} \neq \{\hat{p}_t\}$, such that $\sum_{t=1}^T \big(N_t \cdot \tilde{p}_t \big) = \sum_{t=1}^T \big(N_t \cdot \hat{p}_t \big)$.
Since $N_t>0$ for all $t=1,2,\cdots,T$, there is at least one $k \in \{1,2,\cdots,T\}$ that satisfies $\tilde{p}_k \neq \hat{p}_k$.
If $\tilde{p}_k > \hat{p}_k$, then the same contradiction occurs just like we've discussed above.
If $\tilde{p}_k < \hat{p}_k$, then there must exist another $\tilde{p}_l > \hat{p}_l$ to maintain $\sum_{t=1}^T \big(N_t \cdot \tilde{p}_t \big) = \sum_{t=1}^T \big(N_t \cdot \hat{p}_t \big)$.
Either way, the contradiction is unavoidable, which implies that the optimal pricing strategy $\{\hat{p}_t\}$ is unique.
\end{IEEEproof}

\section{Proof of Proposition 5}\label{sec_Appendix_Pro5}

\begin{IEEEproof}
We use $\phi$ (in radian) instead of $\theta$ (in degree) to denote the elevation angle, where $\phi = \theta \cdot \pi/180^\circ $.
For a user with horizontal distance $r$ to the UAV, the average pathloss is given by $\overline{L}_{UAV}(\phi,r)=L_{LoS}(d) P_{LoS}(\theta) +L_{NLoS}(d) \big[1-P_{LoS}(\theta)\big]$.
With minor deduction, we have
\begin{equation}\label{eqn_PathLossSimplify}
\overline{L}_{UAV}(\phi,r)=L_{NLoS}(d)- \eta \cdot P_{LoS}(\theta),
\end{equation}
where $\eta\!=\!\eta_{NLoS}\!-\!\eta_{LoS}\!<\!\eta_{NLoS}$, $d\!=\!\dfrac{r}{\cos\phi}$, $\theta\! =\! \dfrac{180^\circ}{\pi}\phi$.
By denoting $\eta_{NLoS}$ as $L_1$ and $\eta \cdot P_{LoS}(\theta)$ as $L_2$ to simplify the writing, we can provide the following assertions based on (\ref{eqn_PathLoss}): As $\phi$ increases from $0$ to $\pi/2$, $L_1$ increases monotonously from $L_{NLoS}(r)$ to infinity, while $L_2$ monotonously increases within a sub-interval of $(0,\eta)$.
 Therefore, $0\!<\!\overline{L}_{UAV}(0,r)\!<\!L_{NLoS}(r)$, and $\overline{L}_{UAV}(\phi,r) \!\rightarrow\! +\infty$ as $\phi \!\rightarrow\! \pi/2$.
In addition, $\overline{L}_{UAV}(\phi,r)$ has lower bound, $\big[L_{NLoS}(r)\!-\!\eta_{NLoS}\big]$, in the whole definition domain $[0, \pi/2]$.
By considering the partial derivative of $\overline{L}_{UAV}(\phi,r)$ with respect to $\phi$, we have
\begin{equation}\label{eqn_PathLossDerivative}
\dfrac{\partial \overline{L}_{UAV}(\phi,r)}{\partial \phi}= \dfrac{\partial L_1}{\partial \phi} - \dfrac{\partial L_2}{\partial \phi}=\dfrac{20}{\ln 10} \tan{\phi} - \dfrac{180^\circ\pi^{-1} ab\eta \exp{[-b(\theta-a)]}}{{\{ 1+a\exp{[-b(\theta-a)]} \}^2}}.
\end{equation}
where we have ${\partial L_1}/{\partial \phi}=0$ as $\phi=0$, ${\partial L_1}/{\partial \phi}\rightarrow +\infty$ as $\phi \rightarrow + \infty$, and ${\partial L_2}/{\partial \phi}>0$ as $\forall \phi \in [0, \pi/2)$.
Therefore, we can conclude that $\overline{L}_{UAV}(\phi,r)$ decreases near $\phi=0$ and rapidly increases to $+\infty$ near $\phi=\pi/2$.

By now we have confirmed that: (a) $\overline{L}_{UAV}(\phi,r)$ decreases near $\pi=0$; (b) $\overline{L}_{UAV}(\phi,r)$ increases to infinity as $\phi \rightarrow \pi/2$; and (c) $\overline{L}_{UAV}(\phi,r)$ has a lower bound in $[0,\pi/2)$.
Therefore, there is at least one minimal value as $\phi\in(0,\pi/2)$ that is smaller than $\overline{L}_{UAV}(0,r)$, which makes the existence of a minimum value as $\phi\in(0,\pi/2)$.
Fig.~\ref{fig_Pathloss} provides a exemplary illustration of $\overline{L}_{UAV}(\phi,r)$ with different $r$ values.

\begin{figure}[!thp]
\centering
\includegraphics[width=4.0in]{{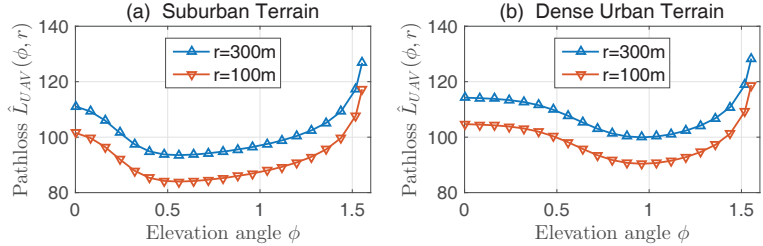}}
\vspace{-5mm}
\caption{(a) shows the pathloss in a typical suburban terrain, where parameters $a=5$, $b=0.2$, $\eta_{LoS}=0.1$, and $\eta_{NLoS}=21$. (b) shows the pathloss in a typical dense urban terrain, where parameters $a=14$, $b=0.12$, $\eta_{LoS}=1.6$, and $\eta_{NLoS}=23$.}\label{fig_Pathloss}
\vspace{-5mm}
\end{figure}

The effective offloading region of the UAV, however, is based on the SNR of each possible location.
Rigorous mathematical analysis would be highly difficult, thus only a simple discussion is provided as following.
Since we have assumed that the UAVs have the same height and the fixed horizontal locations, we can first conclude that, if a user is horizontally nearest to UAV$_n$, then the SNR from UAV$_n$ is always the largest among all the UAVs no matter how large $H$ is.
Therefore, the user partition among UAVs are independent of $H$, and we only have to care about whether the SNR from UAV$_n$ ($\gamma_{UAV_n}$) is greater than the SNR from the MBS ($\gamma_{MBS}$).
For any given location, the scope of $H$ that satisfies $\gamma_{UAV_n} > \gamma_{MBS}$ can be either an empty interval or one or more disjoint intervals (called as the \emph{effective height interval} of this user), depending on the number and the values of the minimal points of $\overline{L}_{UAV}(\phi,r)$.

At the height of $H$, the effective offloading area of UAV$_n$, (given by $S_n$), depends on whether the value of $H$ resides in the the effective height interval of each possible location on the ground.
The theoretical deduction of the optimal height that maximizes $S_n$ is intractable.
However, the existence of such optimal height can be guaranteed, since the effective height intervals are either empty or within $[0,+\infty)$.
\end{IEEEproof}

\section{Proof of Proposition 6}\label{sec_Appendix_Pro6}

\begin{IEEEproof}
For the MBS optimal contract based on $\{\lambda_t\}$, the bandwidth allocation is denoted as $\{w_t\}$ and the corresponding cost of the MBS is denoted as $C\big(\sum{w_t}\big)$.
If we change the types from $\{\lambda_t\}$ to $\{\lambda_t^{\prime}\}$ and assume that the bandwidth allocation remains to be $\{w_t\}$, the cost of the MBS will still be $C\big(\sum{w_t}\big)$.
Since $\lambda_t \le \lambda_t^\prime$, we have $U(\lambda_t,w)<U(\lambda_t^\prime,w)$ according to Proposition~\ref{pro_Monotonicity}.
And based on (\ref{eqn_OptimalPricing}), we can deduce that $p_t$ will be greater, for any $t=1\cdots T$.
Therefore, the sum of prices will gets larger, and the revenue of the MBS will increase from $\hat{R}$ to $\hat{R}_w$.
Note that the above discussion is based on the assumption that $\{w_t\}$ remain the same, which is probably not an optimal bandwidth allocation for $\{\lambda_t^{\prime}\}$.
If we run the algorithm in Section~\ref{sec_Design_Algorithm}, the final revenue $\hat{R}^\prime$ that based on another bandwidth allocation $\{w_t^\prime\}$ will be greater than $\hat{R}_w$.
Therefore, we have $\hat{R} \le \hat{R}_w \le \hat{R}^\prime$.
\end{IEEEproof}

\end{appendices}


\end{document}